\pdfoutput=1
\documentclass{article}%
\usepackage{amsmath}
\usepackage{amsfonts}
\usepackage{amssymb}
\usepackage{graphicx}%
\providecommand{\U}[1]{\protect\rule{.1in}{.1in}}

\begin{document}

\title{\textbf{\ Rowing and the Same-Sum Problem Have Their Moments}}
\author{John D. Barrow{\large \textbf{$\ $}} \vspace{0.3cm}
\and DAMTP, Centre for Mathematical Sciences,
\and Cambridge University,
\and Wilberforce Rd., Cambridge CB3 0WA, UK\\\vspace{0.2cm}}
\maketitle

\begin{abstract}
We consider the optimal positioning of an even number of identical crew
members in a coxless racing boat in order to avoid the presence of a sideways
wiggle as the boat is propelled forwards through the water. We show that the
traditional (alternate port and starboard) rig of racing boats always
possesses an oscillating non-zero transverse moment and associated wiggling
motion.  We show that the problem of finding the zero-moment rigs is related
to a special case of the Subset Sum problem. We find the one (known)
zero-moment rig for a racing Four and show there are four possible such rigs
for a racing Eight, of which only two (the so called 'Italian' and 'German'
rigs) appear to be already known. We also give the 29 zero-moment solutions
for racing Twelves but refrain from explicitly listing the 263 Sixteens and
2724 Twenties which have zero transverse moments. We show that only balanced
boats with crew numbers that are divisible by four can have the zero-moment
property. We also discuss some aspects of unbalanced boats, in which the
number of port and starboard oars are unequal.

\end{abstract}

\section{ \ \ \ \ \ \ \ \bigskip Messing About in Boats}

In this article we are going to consider some simple mathematical properties
of (heavily idealized) coxless rowing crews. Specifically, we are going to
investigate how the placement of the rowers (the 'rig') of the boat has some
simple dynamical consequences for the motion of the boat. The traditional way
of rigging a boat, places alternates rowers pulling oars on the two sides of
the boat, as in Figure 1 for a Four. We will label this traditional rig
configuration of oarsmen as $udud$ where $u$ and $d$ label the two sides of
the boat (up and down as we look at the page but starboard and port when
viewed from inside the boat).%

\begin{figure}
[ptb]
\begin{center}
\includegraphics[
natheight=1.740000in,
natwidth=5.000400in,
height=1.7772in,
width=5.0548in
]%
{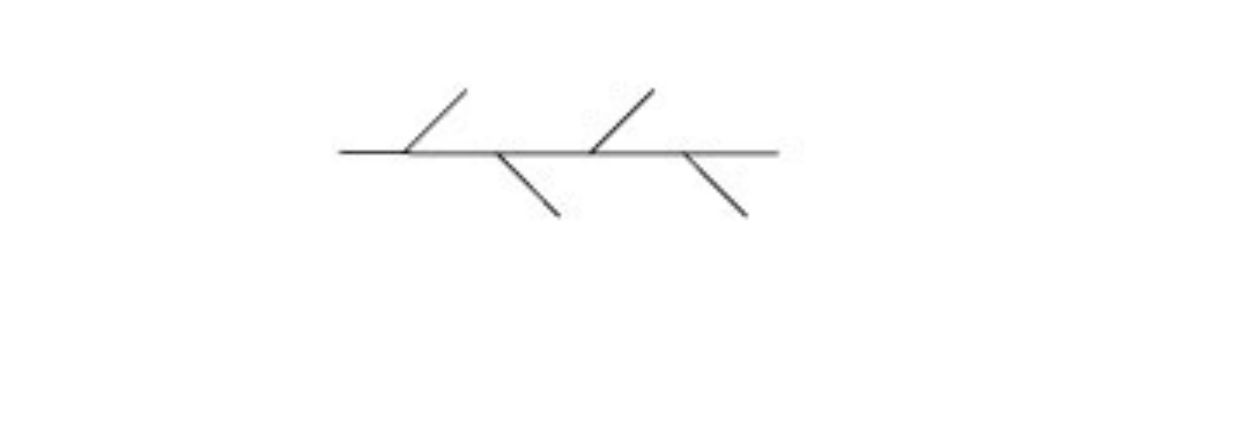}%
\caption[The traditional rig of a Four]{The traditional rig of a Four, which
we label $udud$.}%
\label{fig 1}%
\end{center}
\end{figure}

If we look at the boat from the other side then it will have the
\textit{mirror }configuration $dudu$. We regard these as identical and do not
distinguish rigs under the relabelling transformation
$(u,d)\longleftrightarrow(d,u).$

For simplicity we are going to make some assumptions. All our rowers are
assumed to be identical in strength and technique and to employ identical
equipment with identical set up. They will each pull only a single oar on one
side of the boat (no double scullers). We will not include a cox or any other
influence (rowers of unequal strengths) that might counteract sideways forces.
All of these simplifications can be dropped if necessary \ and adapted
versions of our conclusions will hold. Finally, unless we say otherwise we
will always we considering crews with the same number of oars pulling on each
side of the boat. We will call such a boat a \textit{balanced} boat.

The traditional rig appears symmetrical and simple in ways that might tempt
you into thinking it is in every sense optimal. However, this is not the case,
as we shall now see.

\section{The Moment Problem}

First, consider the set-up of a balanced Four. The standard rig for this boat
has a simple alternation of rowers which we shall call the \textit{standard
rig, }shown in Figure 1. A time-varying force is exerted on the boat by the
torque applied by each rower's oar in its rowlock. The time variation of this
'rowing force curve' is much studied by coaches seeking to balance crews,
improve technique, and understand the strengths and weaknesses of crew members
individually and collectively \cite{force}. We shall ignore all this detailed
analysis together with details of blade shape, angle to the water, flexure and
hydrodynamics \cite{water}, and note simply that the force can always be
resolved into its two perpendicular components: one in the direction of the
boat's forward motion and the other at right angles to it. We shall call the
latter normal component to the line through the stern and the bow the
\textit{transverse} force.%

\begin{figure}
[ptb]
\begin{center}
\includegraphics[
natheight=2.120500in,
natwidth=3.666800in,
height=2.1594in,
width=3.7144in
]%
{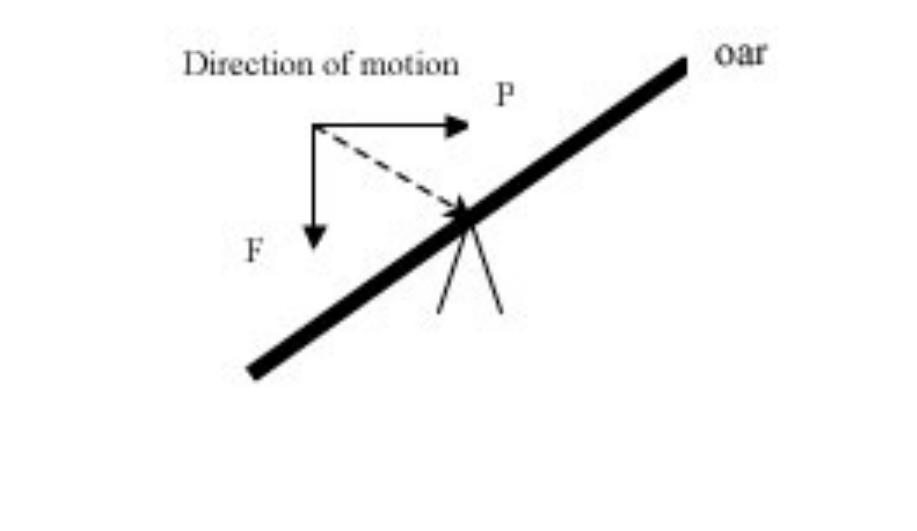}%
\caption[Forces in phase 1]{The orthogonal components of the force exerted on
the boat at the oarlock by an oar. During the first phase of the stroke the
component normal to the direction of motion, $F$, is directed towards the
boat. The component in the direction of motion is $P$.}%
\label{fig 2}%
\end{center}
\end{figure}
%

\begin{figure}
[ptb]
\begin{center}
\includegraphics[
natheight=2.496700in,
natwidth=3.681500in,
height=2.5374in,
width=3.7291in
]%
{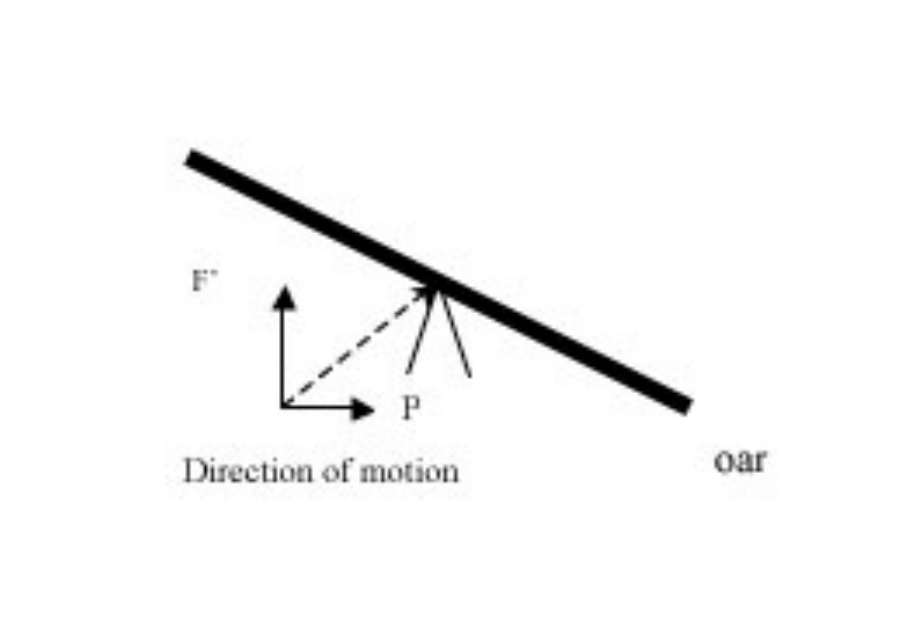}%
\caption[Forces in phase 2]{During the second (recovery) phase of the stroke
the component of the force normal to the direction of motion, $F^{\prime}$, is
reversed and points away from the boat. The component in the direction of
motion, $P$, keeps the same sign.}%
\label{fig 3}%
\end{center}
\end{figure}
\ \ 

During the first (catch and drive) part of the rower's stroke the transverse
force at the oarlock is directed towards the boat (see Figure 2) but during
the second (extraction and recovery) phase of the stroke it reverses and is
directed away from the boat (see Figure 3) \cite{town}. This alternation in
the direction of the transverse component of the force exerted on the boat
continues almost periodically as the rowers repeat their sequence of stroke
patterns. \ \ \ \ \ 

Let us now sum the moments of these transverse forces about any point on the
centre line of the boat and we will pick that point to be the stern. Assume
the distance from the stern to the first rower (the stroke) is $s$ and the
spacing between all the other three rowers in the Four is $x$. During the
first phase of their synchronised strokes assume that each rower exerts the
same transverse force of magnitude $F>0$. The net sideways moment on the
traditionally rigged $udud$ boat is equal to%

\[
M(phase\text{ }1)=sF-(s+x)F+(s+2x)F-(s+3x)F=-2Fx<0
\]
where we have picked the 'up direction in our diagram for the + sign.

There are three simple points to notice from this result for the traditionally
rigged crew of the Four. First, because our crews will always be balanced,
with the same number of oars on each side, the dependence on the distance to
the stroke's location, $s,$ always cancels out and we can set it to any value
we choose and henceforth we set $s=1$. Second, the value of $x$ can also be
set equal to unity by absorbing it into $F$ or $F^{\prime}$ by a redefinition.
Third, and most important, the net moment on the boat during the first phase
of the stroke with the traditional rig is \textit{non-zero}. During the second
recovery phase of the stroke the sign (and possibly the value) of the
transverse force will change to $-F^{\prime}<0$ and the net moment on the boat
will now be%

\[
M(phase\text{ }2)=+2F^{\prime}x>0
\]
As a result, the boat will wiggle steadily from side to side as it moves
forward under the influence of this alternating transverse force. It can be
countered by the rowers (or by a cox) sensing its existence and making small
adjustments. But this takes extra energy from the rowers and slows the forward
progress of the boat. Ideally, we might like to eliminate this transverse
alternating torque on the boat. This wiggle will arise whenever there is an
asymmetry in the force applied by a rower during the whole stroke cycle, even
if there is no change of sign in the normal component. In fact, in our model,
whenever $F\neq F^{\prime}$ there we will be a sideways drift of the boat in
addition to a wiggle. For simplicity, we shall take $F=F^{\prime}$ in what
follows. Although this is not entirely realistic, it does not affect our
results because we are going to be interested in finding situations where the
net moment on the boat vanishes. From now on we will also set $x\equiv s\equiv
F\equiv F^{\prime}\equiv1.$

Let's stay with the possible balanced rigs for the Four. There are only three
distinct possibilities, shown in Figure 4.%

\begin{figure}
[ptb]
\begin{center}
\includegraphics[
natheight=2.629000in,
natwidth=3.884700in,
height=2.6714in,
width=3.9332in
]%
{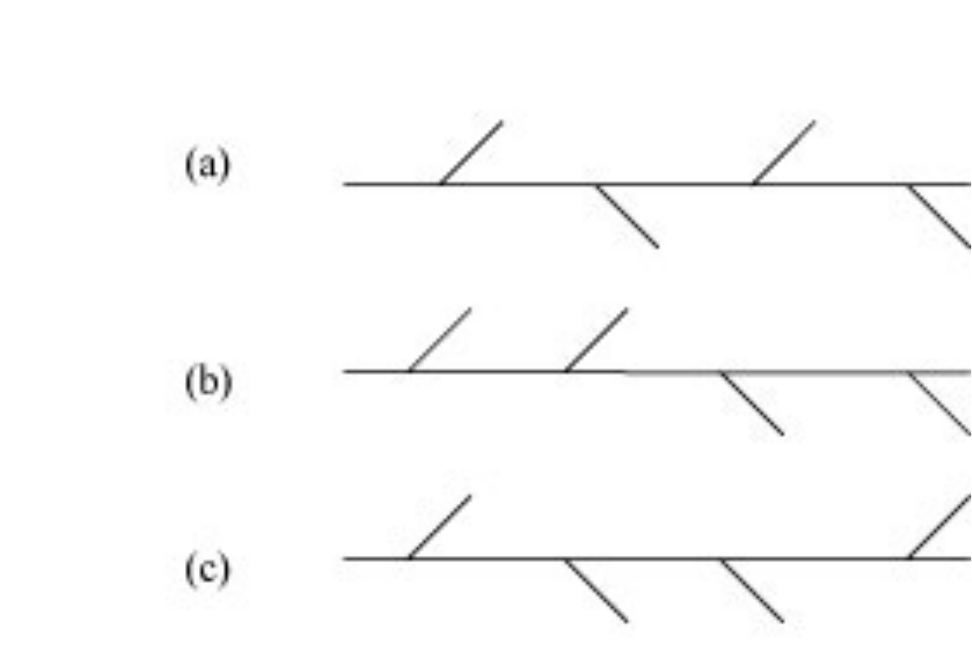}%
\caption[Three Fours]{The three distinct ways to rig a Four: (a) The
traditional rig, $udud$, with transverse moment $\pm2$; (b) The worst rig,
$uudd$, with transverse moment $\pm4$; (c) The 'Italian' rig, $uddu$, with
zero transverse moment.}%
\label{fig 4}%
\end{center}
\end{figure}

We can swop the $u$ and $d$ positions but that is just a mirror of the
original (eg $udud$ and $dudu$ are both the standard rig) and we don't count.

The standard rig (a) has a non-zero moment and a wiggle with $M=\pm2$. The rig
(b) is the worst way to rig a Four, with a transverse moment that is twice
that of the standard rig:%

\[
\pm M=1+2-3-4=-4.
\]

However, the last possibility, (c), has a zero moment ,%

\begin{equation}
\pm M=1-2-3+4=0 \label{b}%
\end{equation}

and for this configuration there will be no wiggle in the boat's forward
motion. This configuration is known as the 'Italian' rig because it was
discovered by the Moto Guzzi Club team on Lake Como in 1956. The crew of the
Club's Four was being watched by Giulio Cesare Carcano, one of the company's
leading motorcycle engineers from Milan, who suggested that the failure of the
boat to run straight might be alleviated by putting the middle two oarsmen
both on the starboard side with the stroke and the bow still on the port side
\cite{and}. The result was rather successful and the Moto Guzzi crew went on
to represent Italy and take the Gold Medal that year at the Melbourne Olympic Games.

\section{A 'Momentous' General Formulation}

We can see from the simple example of the Four that the values the transverse
moments during each of the two phases of the stroke are given by%

\begin{align*}
(a)  &  :udud:1-2+3-4=-2\\
(b)  &  :uudd:1+2-3-4=-4\\
(c)  &  :uddu:1-2-3+4=0
\end{align*}
The zero-moment rig is the one in which the sum of the first 4 integers can be
combined with 2 plus signs and 2 minus signs, as required for balance, so that
their sum is zero. The first entry, +$1$, will always be taken positive
without any loss of generality because the case with $-1$ will just produce
the mirror rig.

This formulation allows us to look at the generalisations to balanced crews
containing with zero transverse moment with any even number of rowers. If
there are $2N$ rowers, with $N=1,2,3,...$ then we are looking for the set of
solutions to the arithmetic problem%

\begin{equation}
1\pm2\pm3\pm.......\pm2N=0 \label{a}%
\end{equation}
in which the left-hand side contains $N$ plus signs and $N$ minus signs.

We can immediately draw a simple conclusion by using this formulation:%

\[
\text{\textit{Balanced} \textit{zero-moment rigs are only possible with 2N
rowers if }N\textit{\ is even};}
\]
that is, the number in the crew must be divisible by $4.$ Thus, there cannot
be a balanced Six with zero transverse moment. The proof is simple: we have to
add or subtract $N$ even numbers and $N$ odd numbers to get zero. This is only
possible if $N$ is even because only then will the combination of the $N $ odd
numbers be even like the combination of the $N$ even numbers and only then can
the sum of both be zero. The simplest case of all, that of a pair ($N=1$) also
always has a non-zero moment equal to $1-2=-1$, or $+1=-1+2 $ for the mirror
rig. Hence, we can now confine our attention to crews of 4, 8, 12, etc.

We can streamline the formulation in a further interesting fashion by noting
that the sum requirement (\ref{a}) for a zero moment is equivalent to the
requirement that the sum of all the numbers from $1$ to $2N$ is equal to twice
the sum of all those entering with + signs (or minus all those entering with
minus signs):%

\[
\sum_{r=1}^{2N}r+2\left(  \sum\operatorname{negative}entries\right)
=0=\sum_{r=1}^{2N}r-2\left(  \sum\operatorname{positive}entries\right)
\]
where the second sum is just over the $N$ entries with negative (or positive)
signs, respectively. This means, for a crew of $2N$ rowers,%

\[
N(2N+1)+2\left(  \sum\operatorname{negative}entries\right)
=0=N(2N+1)-2\left(  \sum\operatorname{positive}entries\right)  .
\]
So we just need to find the cases where the sum of the $N$ negative (or
positive) entries add up to%

\begin{equation}
\ \frac{N(2N+1)}{2} \label{d}%
\end{equation}
For the case of the Four ($N=2$) we need the sum of the $N=2$ negative entries
to be equal to -5, which was indeed the case for the zero-moment configuration
(c) given above in (\ref{b}).

This problem is now recognisable as a version of the \textit{Subset-Sum
problem} in which a subset, $S_{k}$ of a set positive integers, $S$, is
selected by the requirement that the sum of the members of the subset be equal
to some integer, $k$. The version of this algorithmic problem that we are
faced with here is the \textit{Same Sum problem} \cite{stan} in which we want
to find all the subsets $S_{k}$ of $S$ which have the same sum, $k=N(2N+1)/2$%
.  Jeffrey Shallit has also pointed out a connection to structure of the
Thue-Morse sequence \cite{shall}.

\section{Eights, and After}

As with the Four, the traditional rig for a rowing Eight has a significant
non-zero moment ($M=\pm4$) and a counter-productive transverse wiggle for the
cox to counter. This non-zero moment corresponds to a non-zero sum in which
the sum of the negative entries is $-20$ rather than the value $-18$ required
for a non-zero moment because:%

\[
udududud:1-2+3-4+5-6+7-8=-4
\]
For the case of a rowing Eight, where $N=4$, the Same Sum problem for a
zero-moment rig is \ the case where the sum of the magnitudes of the 4
negative entries the alternating sum must equal $-18$. There are four distinct
solutions to this problem:%

\begin{align*}
(a)  &  :3+4+5+6=18\text{ ie }uudddduu\\
(b)  &  :2+4+5+7=18\text{ ie }ududdudu\\
(c)  &  :2+3+6+7=18\text{ ie }udduuddu\\
(d)  &  :2+3+5+8=18\text{ ie }udduduud
\end{align*}

\bigskip

These (together with their four mirrors) give the four possible zero-moment
rigs for an Eight shown in Figure 5.%

\begin{figure}
[ptb]
\begin{center}
\includegraphics[
natheight=3.255200in,
natwidth=3.872600in,
height=2.412in,
width=2.8634in
]%
{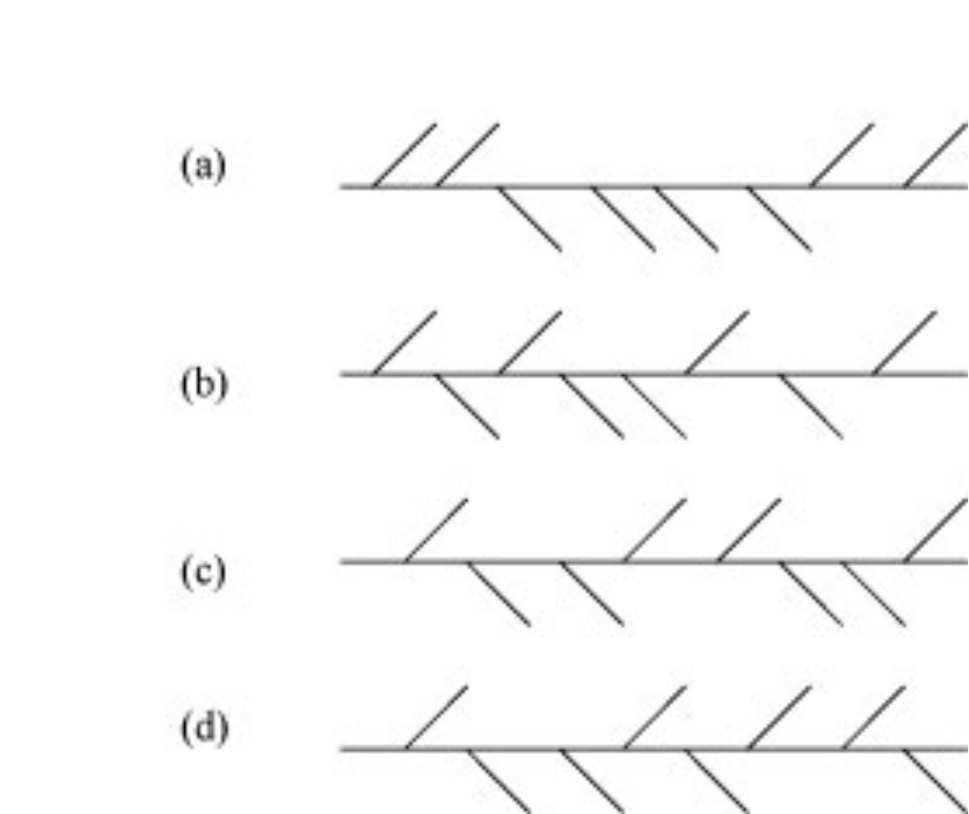}%
\caption[Four Eights]{ The four possible rigs for Eights which have zero
transverse moment. (a) The New rig 1: $uudddduu$; (b) The `German' rig:
$ududdudu$; (c) The `Italian' rig: $udduuddu $; and (d) The New rig 2:
$udduduud.$}%
\label{fig 5}%
\end{center}
\end{figure}

Two of these are known. Rig (c) is the so called 'Italian' or 'triple tandem'
rig, that was used by the Italian eights in the 1950s as an extension of
Carcano's insight about the Four by adding two zero-moment ($uddu$) Fours
together in series. Rig (b) is the so called 'German', 'bucket', or 'Ratzeburg
rig', first used by crews training at that famous German rowing club in the
late 1950s under Karl Adam, who was motivated by Carcano's configuration. Both
the German and Italian crews were successful with these two zero-moment Eights
and the zero-moment Four at the 1958 European Championships and the German
crews coached by Adam went on to be supremely successful over the next ten years.

The other two zero-moment rigs found here, (a) and (d), appear to be new and
have not previously been discussed. Note that (d) is a combination of a
zero-moment Italian Four with a mirrored zero-moment Italian Four. The other
new rig, (a) is special because it manages to have a quadruple tandem
configuration with a same-side Four positioned inside two same-side pairs.

\bigskip The zero-moment eights have \ simple construction from the three
possible Fours. All zero-moment Eights can be made by adding Fours with
moments $-4+4,-2+2,$ $0+0,$ and $0-0$, where the minus sign indicates the
switch $u\leftrightarrow d$ $.$ These four combinations correspond to the
solutions (a), (b), (c) and (d), above respectively.

If we increase the number of rowers the number of permutations grows quickly.
There are 29 distinct zero-moment rigs for 12-person crews, 263 zero-moment
16-person crews and 2724 zero-moment 20-person crews. For example, all the
12-rower crews correspond to all the solutions of the Same Sum problem,
eq.(\ref{d}), for which six of the numbers in $\{1,2,3...,12\}$ sum to 39.
Here are all of these 29 solutions for the 12 seater crews \cite{sean}. The
numbers appearing in each sum give the positions of the $d$ rowers :%

\begin{align*}
39  &  =4+5+6+7+8+9=3+5+6+7+8+10=3+4+6+7+9+10\\
&  =2+5+6+7+9+10=3+4+5+8+9+10\ \ \\
39  &  =2+4+6+8+9+10=1+5+6+8+9+10=2+3+7+8+9+10\\
&  =1+4+7+8+9+10=3+4+6+7+8+11\\
39  &  =2+5+6+7+8+11=3+4+5+7+9+11=2+4+6+7+9+11\\
&  =1+5+6+7+9+11=2+4+5+8+9+11\\
39  &  =2+3+6+8+9+11=1+4+6+8+9+11=1+3+7+8+9+11\\
&  =3+4+5+6+10+11=2+4+5+7+10+11\\
39  &  =2+3+6+7+10+11=1+4+6+7+10+11=2+3+5+8+10+11\\
&  =1+4+5+8+10+11=1+3+6+8+10+11\\
39  &  =1+2+7+8+10+11=2+3+4+9+10+11=1+3+5+9+10+11\\
&  =1+2+6+9+10+11.
\end{align*}

Each, as always, has a mirror solution in which these numbers give the
positions of the $u$ crew members. For example, the mirror of the first
solution given in the list corresponds to $1+2+3+10+11+12=0.$ Again, the form
of these zero-moment solutions can be analysed as composites of Eights and
Fours with moments that combine so as to sum to zero in all possible ways.

\section{Unbalanced boats}

We have ignored the situation of unbalanced boats so far. They have unequal
numbers of oars in the water on the two sides of the boat and so there will
always be an asymmetry in the total forward force on the boat on each side and
it will veer away from its straight-line forward course unless corrective
action is taken. However, it is possible to have a zero transverse moment in
these cases under particular conditions. Because the boat is no longer
balanced, the size of $s$, the distance from the bow to the first rower will
matter. Consider a Three with the rower's stations at distances $s,s+x,s+2x$
from the stern. If $s=x,$ there is a rig ($uud$) with zero transverse moment
(shown in Figure 6) since the total moment is now%

\[
M=sF+(s+x)F-(s+2x)F=(s-x)F
\]
%

\begin{figure}
[ptb]
\begin{center}
\includegraphics[
natheight=1.004900in,
natwidth=2.499300in,
height=1.0378in,
width=2.5399in
]%
{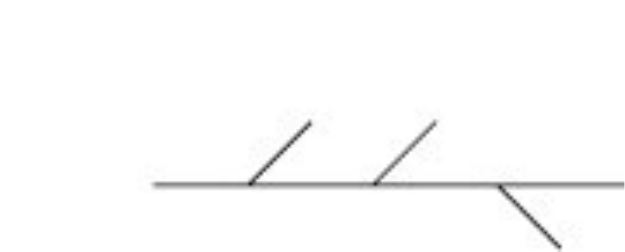}%
\caption[Unbalanced Three]{An unbalanced Three with $x=s.$}%
\label{fig 6}%
\end{center}
\end{figure}

Similarly, as shown in Figure 7, we can make an unbalanced Six with a $uuudud
$ configuration that has moment%

\[
M=sF+(s+x)F+(s+2x)F-(s+3x)F+(s+4x)F-(s+5x)F=(2s-x)F
\]
which will be zero is we pick $x=2s$. This is a general feature of unbalanced
zero-transverse moment boats. The spacing of the stroke and bow position from
the ends of the boat is constrained by the separation of the other rowers.%

\begin{figure}
[ptb]
\begin{center}
\includegraphics[
natheight=1.004900in,
natwidth=2.386000in,
height=1.0378in,
width=2.4267in
]%
{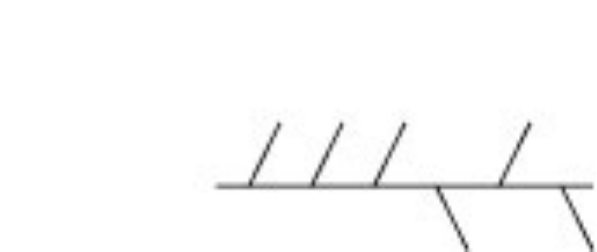}%
\caption[Unbalanced Six]{An unbalanced Six $uuudud$ where $x=\ 2s.$}%
\label{ fig 7}%
\end{center}
\end{figure}

\section{Conclusions (and Apologies)}

Our idealized discussion of the transverse moments exerted by the crews of
racing shells is not really intended to revolutionise the tactics adopted by
world-class racing crews (although we would be happy for that to be the case
none the less). If it provides an instructive example of some simple
mechanical principles in conjunction with an illustration of how a
combinatorial problem arises in a familiar sporting context then it will have
been sufficiently 'momentous'.

\textbf{Acknowledgements }I would like to thank Bill Atkinson, Baojiu Li, Sean
Lip and Owen Smith for their assistance, and James Cranch and Jeffrey Shallit
for helpful communications and references.

\bigskip

\bigskip

\bigskip

\bigskip

\bigskip

\bigskip

\bigskip

\bigskip

\bigskip

\bigskip

\bigskip

\bigskip

\bigskip

\bigskip

\bigskip

\bigskip

\bigskip

\bigskip

\bigskip

\bigskip

\bigskip

\bigskip

\bigskip

\bigskip

\bigskip

\bigskip

\bigskip

\bigskip

\bigskip

\bigskip

\bigskip

\bigskip

\bigskip

\bigskip


\begin{thebibliography}{9}
\bibitem {sean}I am indebted to Sean Lip, for generating this list as well as
its counterpart for larger boat crews which we do not include here because of
limitations of space.
\end{thebibliography}
\end{document}